\documentclass[useAMS,usenatbib]{mn2e}

\usepackage{graphicx}
\usepackage{subfigure}
\usepackage{natbib}

\def\fp{\hbox{$.\!\!^{\reset@font\scriptscriptstyle\r@mn{p}}$}}

\long\def\crap#1{}

\def\<{\langle }
\def\>{\rangle }

\title[The IMFs of M31 and M32]{The initial mass functions of M31 and M32 through far red stellar absorption features}
\author[Zieleniewski et al.]{Simon Zieleniewski$^{1}$\thanks{E-mail:
simon.zieleniewski@physics.ox.ac.uk} Ryan C. W. Houghton$^{1}$ Niranjan Thatte$^{1}$ Roger L. Davies$^{1}$\\
$^{1}$Astrophysics, Denys Wilkinson Building, Keble Road, Oxford, OX1 4RH}
\begin{document}

\date{Submitted 14/02/15}

\pagerange{\pageref{firstpage}--\pageref{lastpage}} \pubyear{2014}

\maketitle

\label{firstpage}

\begin{abstract}
Using the Oxford Short Wavelength Integral Field specTrograph, we investigate radial variations of several initial mass function (IMF) dependent absorption features in M31 and M32. We obtain high signal-to-noise spectra at six pointings along the major axis of M31 out to $\sim700''\,(2.7\,\rmn{kpc})$ and a single pointing of the central $10\,\rmn{pc}$ for M32. In M31 the sodium NaI $\lambda$8190 index shows a flat equivalent width profile at $\sim0.4\,\rmn{\AA}$ through the majority of the bulge, with a strong gradient up to $0.8\,\rmn{\AA}$ in the central $10'' (38\,\rmn{pc})$; the Wing-Ford FeH $\lambda$9916 index is measured to be constant at $0.4\,\rmn{\AA}$ for all radii; and calcium triplet CaT $\lambda$8498, 8542, 8662 shows a gradual increase through the bulge towards the centre. M32 displays flat profiles for all three indices, with FeH at $\sim0.5\,\rmn{\AA}$, very high CaT at $\sim0.8\,\rmn{\AA}$ and low NaI at $\sim0.1\,\rmn{\AA}$. We analyse these data using stellar population models. We find that M31 is well described on all scales by a Chabrier IMF, with a gradient in sodium enhancement of [Na/Fe]$\,\sim+0.3\,\rmn{dex}$ in the outer bulge, rising within the central $10''$ to perhaps [Na/Fe]$\,\sim+1.0\,\rmn{dex}$ in the nuclear region. We find M32 is described by a Chabrier IMF and young stellar age in line with other studies. Models show that CaT is much more sensitive to metallicity and [$\alpha$/Fe] than to IMF. We note that the centres of M31 and M32 have very high stellar densities and yet we measure Chabrier IMFs in these regions.
\end{abstract}


\section{Introduction}

One of the most rigorous methods of constraining galaxy formation and evolution is through analysis of their stellar spectra. Stars are the directly observable baryonic component of galaxies, which habour information about their formation. Integrated light from the unresolved stellar component of galaxies thus presents a direct probe of their composition. The initial mass function (IMF) is one of the most important parameters to constrain our understanding of galaxy formation. This function sets the mass distribution in stars and affects the total mass-to-light ratio of the galaxy. It has been one of the most difficult functions to constrain for external galaxies, as direct star counts are not possible. The IMF is defined as the number density of stars $n$ per unit mass $m$ formed at one time. The work of \citet{Salpeter1955} \citep[and later][]{Scalo1986}  showed that the IMF $\xi (m)$ takes a simple functional form in the Milky Way
\begin{equation}
\xi (m) = \frac{\rmn{d}n}{\rmn{d}m} = km^{-x},
\end{equation}
and the exponent $x$ takes a value of $x=2.35$, down to masses of $\sim 1\,\rmn{M_{\odot}}$. The low mass end of the IMF is especially challenging to constrain from integrated light, as stars of mass $< 0.5\,\rmn{M}_{\odot}$ contribute only $\sim1$ per cent of the bolometric luminosity of the galaxy. However, several studies extended the form of the IMF below $1\,\rmn{M}_{\odot}$, and suggested either a flattening of the slope \citep{MillerScalo1979} or even a turnover around $1\,\rmn{M}_{\odot}$ \citep{Kroupa2001, Chabrier2003}.

It has long been assumed that the form of the IMF is universal for all galaxy types, masses and ages. Indeed, most galaxy formation and large scale structure simulations simply invoke a single form of the IMF for all simulations. Throughout the second half of the 20th century, several papers highlighted and detected specific stellar absorption features that are prominent in the spectra of low mass stars, but absent in high mass counterparts (e.g. Wing-Ford band FeH $\lambda$9916: \citealt{WingFord1969}, \citealt{Schiavon1997b}, \citealt{Cushing2003}; sodium NaI $\lambda$8190: \citealt{FaberFrench1980}, \citealt{Schiavon1997a}). A lack of both high signal-to-noise and detailed stellar populations models, coupled with the challenges of spectroscopy in the far red and near-infrared, meant that arguments for a varying IMF were inconclusive. However, recent progress by \citet{VanDokkumConroy2010} and \citet{ConroyVanDokkum2012b} has shown that it is possible to detect and model these absorption features in integrated light from massive early type galaxies (ETGs) with very high signal-to-noise spectra. They detected the NaI doublet feature at $8190\,\rmn{\AA}$, the calcium triplet CaT around $8500\,\rmn{\AA}$, and the FeH Wing-Ford band at $9916\,\rmn{\AA}$ in spectra from the centres of 38 giant ETGs. Their results have shown that these galaxies are well modelled by bottom-heavy ($x>2.35$) IMFs and that the IMF becomes more bottom-heavy with increased velocity dispersion. The complimentary approaches of lensing/dynamical modelling to constrain the form of the IMF in ETGs (e.g. strong lensing; \citealt{Treu2010, Smith2015}, and integral field kinematics; \citealt{Cappellari2012}), has also provided consistent constraints with spectroscopic methods \citep[although quantitive differences exist; see][]{Smith2014}. Further, work by \citet{Geha2013} using HST deep imaging has shown evidence for a very bottom-light IMF in very diffuse galaxies, lending further weight to non-universality of the IMF.

While the work by \citeauthor{ConroyVanDokkum2012a} and others shows a paradigm shift is required from the assumption of a constant IMF for all galaxies, their results are only representative of the central regions of each galaxy. Searching for radial gradients in these absorption features and the IMF will provide more conclusive results for use in galaxy modelling and stellar population analysis. Recent results by \citet{Martin-Navarro2015} have indicated a gradient to the IMF slope in a massive early type galaxy with large central velocity dispersion. However, work by \citet{SmithLucey2013} indicated the presence of massive ETGs with bottom-light IMFs. These results show that our knowledge of the IMF is still uncertain.

While all of the recent IMF work has looked at ETGs, it would be interesting to compare to the bulges of spiral galaxies. M31 is an ideal target for high S/N resolved spectroscopy due to its proximity. It is an archetypal spiral galaxy and the central bulge has had its stellar populations extensively studied, both using spectroscopy \citep[e.g.][]{Davidge1997, Saglia2010} and photometry \citep[e.g.][]{Stephens2003, SarajediniJablonka2005}. Determining the presence of IMF gradients will help provide a more comprehensive understanding of spiral bulges and the process behind IMF variations.

The form of the IMF in the nuclear region of M31 has been the subject of much debate throughout the second half of the 20th century \citep[see][section 4.2]{ConroyVanDokkum2012b}. The core has been known to have prominent sodium absorption and this has been interpreted as an increased dwarf abundance by some \citep[e.g.][]{FaberFrench1980} and sodium enhancement by others \citep[e.g.][]{AlloinBica1989}.

In this paper we present three studies on:\\
 a) the variations of far red absorption features in M31 through the bulge out to the old disc at $\sim 700''\,(2.7\,\rmn{kpc})$;\\
 b) the variations of these features through the inner bulge to the nuclear region of M31 where there has been much historic debate;\\
 c) these features in the central $8''\,(30\,\rmn{pc})$ of M32.
 
 We use the definitions from \citet{KormendyBender1999} and \citet{Tempel2011} to describe the components of M31. The nuclear region is defined as $r<4'' (15\,\rmn{pc})$, the bulge as $4'' < r < 450'' (1.7\,\rmn{kpc})$, and the disc as $r > 450''$. We also refer to the region between $4'' < r < 10''$ as the inner bulge.
 
 We analyse our results using the latest stellar population synthesis models from \citet[][hereafter V12]{Vazdekis2012} and \citet[][hereafter CvD12]{ConroyVanDokkum2012a}. We further investigate a method of disentangling the degeneracies in age and metallicity by using colour as a proxy for age, in the absence of the standard H$\beta$ absorption line.

This paper is organised as follows: section 2 details our observations and data reduction procedures. In section 3 we present the absorption feature strengths, and section 4 contains our stellar population analysis using simple stellar population (SSP) models and colour-index maps. We discuss our results in section 5 and conclude our work in section 6. In this paper we adopt a distance to M31 of $785\,\rmn{kpc}$ as measured by \citet{McConnachie2005}, which gives a scale of $3.8\,\rmn{pc}\,\rmn{arcsec}^{-1}$. We further adopt the IMF naming convention used by CvD12, namely a Chabrier IMF corresponding to that defined in Chabrier (2003) for the disc of the Milky Way; a Salpeter IMF defined as $x=2.3$ for V12 and $x=2.35$ for CvD12; and a bottom-light IMF as defined in \citet{VanDokkum2008}.

\section[]{Observations and data reduction}

\subsection{Overview}

\begin{figure*}
 \centering
 \includegraphics[width=12cm]{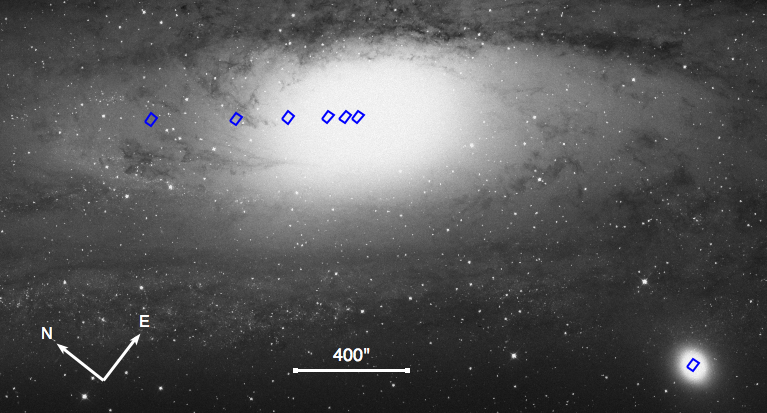}
 \caption{STScI/Digital Sky Survey (DSS) image of M31 and M32 showing the locations of the SWIFT fields (blue boxes). The M31 fields are named in order of increasing radius from M31\_1 through to M31\_6. Note that the boxes only show the locations of the fields, and are not representative of the size of the SWIFT field of view.}
 \label{fig:fields}
\end{figure*}

Observations of M31 and M32 were obtained over the nights of 3rd and 5th December 2011; 23rd and 24th September 2012; and 2nd October 2014 using the Short Wavelength Integral Field specTrograph (SWIFT) instrument \citep{Thatte2006} on the Palomar $200$ inch (5.1 m) telescope. Fig.~\ref{fig:fields} shows the position of each pointing. Observations were taken at the $235$ mas spaxel$^{-1}$ spatial scale covering a field of $10'' \times 21''$. The spectra cover the wavelength range $6300-10400\,\rmn{\AA}$ with a dispersion of $\sim 2\,\rmn{\AA}$ FWHM and a sampling of $1\,\rmn{\AA\,pix^{-1}}$. Table~\ref{tab:obs_fields} lists the pointings and dates of the observations. The seeing was around $1.5''$ for all observations.

\begin{table}
   \caption{Observational fields of M31 and M32. 
  Observations of M31 are labelled as M31\_1 through 
  to M31\_6 in increasing radii from the centre.
  We also give the positions of the sky fields.}
  \label{tab:obs_fields}
  \begin{tabular}{@{}lccc@{}}
  \hline
Field & RA & DEC & Date\\
 \hline
 M31\_1 & 00:42:44.57 & $+$41:16:05.7 & 03-12-2011\\
 M31\_2 & 00:42:46.82 & $+$41:16:41.3 & 23-09-2012\\
 M31\_3 & 00:42:51.04 & $+$41:17:38.3 & 23-09-2012\\
  M31\_4 & 00:42:58.19 & $+$41:19:17.8 & 02-10-2014\\
 M31\_5 & 00:43:10.80 & $+$41:22:03.7 & 23-09-2012\\
 M31\_6 & 00:43:28.30 & $+$41:25:52.8 & 23-09-2012\\
 M32 & 00:42:41.86 & $+$40:51:57.0 & 23-09-2012\\
 SKY\_M31 & 00:44:12.04 & $+$40:48:05.6 & \\
 SKY\_M32 & 00:46:15.27 & $+$41:01:28.3 & \\
\hline
\end{tabular}
\end{table}

The data were reduced using the SWIFT data reduction pipeline, written in {\sc iraf} (Houghton, in preparation). The pipeline handles all the standard reduction processes of bias subtraction, flat fielding, wavelength calibration, error propagation as well as IFS specific features of illumination correction and cube reconstruction. Cosmic rays were detected and removed using the {\sc lacosmic} routine \citep{VanDokkum2001}.

We obtained Sloan Digital Sky Survey (SDSS) $g-r$ colour data from \citet{Saglia2010} and \citet{Tempel2011}. \citeauthor{Saglia2010} covered all M31 pointings except M31\_6 for which we used \citet{Tempel2011}. Sodium D (NaD $\lambda$5893) index measurements for M31 and M32 were obtained from \citet{Davidge1991, Davidge1997}. We applied corrections to the \citeauthor{Davidge1997} NaD data based upon the CvD12 models to account for the measurements being at a lower resolution than $\sigma=200\,\rmn{km}\,\rmn{s}^{-1}$ (see section 2.5).

\subsection{Sky subtraction and spectra extraction}

Sky subtraction was performed to first order by subtracting sky frames observed adjacent in time to each science frame. We were careful to choose sky fields sufficiently far away as to avoid contamination from M31. The positions are given in table~\ref{tab:obs_fields}. We extracted spectra in two ways.

For fields M31\_2 to M31\_6 we summed up all spaxels within the field to obtain a median spectrum of maximal S/N; in this case we first normalised by the median image to remove the increased weighting of brighter regions over fainter ones.

For M31\_1 and M32 fields we performed sector binning \citep{Houghton2013} to obtain a 2D map of spectra with S/N$\,> 100\,\rmn{pixel}^{-1}$. This gave $\sim200$ individual spectra within a radius of $10''$ for M31\_1 and $\sim170$ spectra within $8''$ for M32. The numbers are slightly lower for the FeH optimised spectra due to more demanding S/N requirements for that region of the spectrum.

\subsection{Telluric correction}

Atmospheric telluric absorption is prominent in both the NaI and FeH regions. We observed A0V stars during the observations to act as telluric standards. These stars have strong Paschen lines around 8500$\,\rmn{\AA}$ which we remove using a dedicated routine written in {\sc idl}. The telluric spectra are divided out by model A0V spectra provided by R. Kurucz, using a version of the amoeba algorithm allowing for velocity shifts and stellar rotation. Our corrected telluric spectra are accurate to around the $0.5 - 1$ per cent level (see Fig.~\ref{telluric_correction} in appendix).

\begin{figure*}
 \centering
 \includegraphics[width=17.5cm]{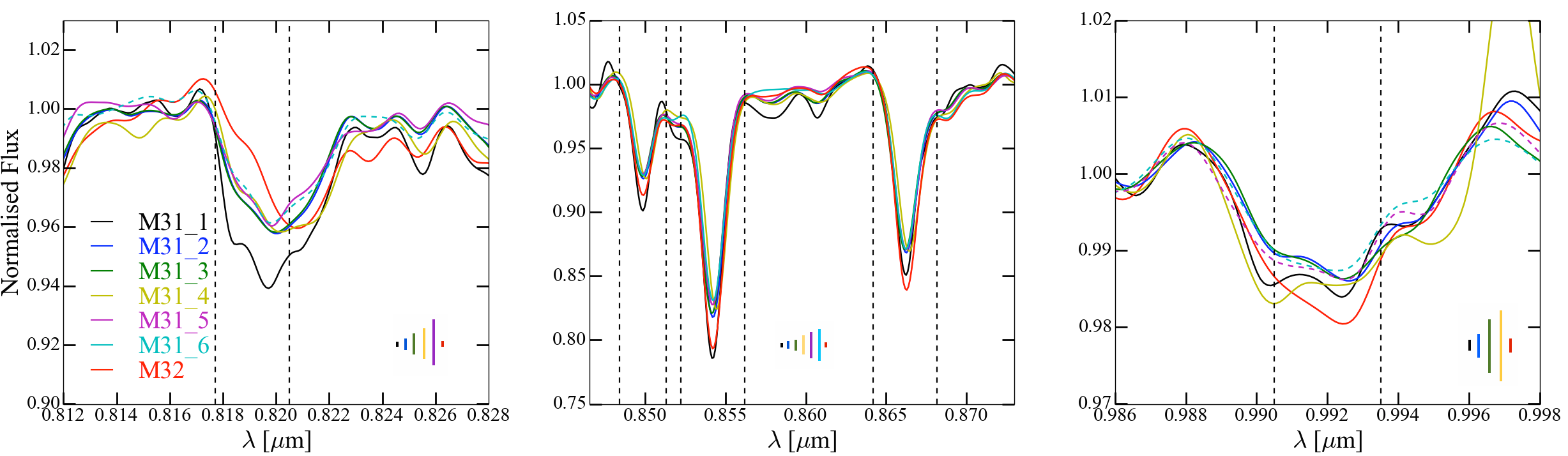}
 \caption{Plots of the three IMF sensitive spectral features, NaI (left), CaT (centre) and FeH (right) for each pointing in M31 and M32. The feature definitions (table~\ref{indices}) are shown by the vertical dashed lines in each subplot. The spectra have been normalised to the continuum bands either side of the feature. M31\_1 and M32 correspond to median spectra obtained over the entire M31\_1 and M32 fields. For M31\_6 in the NaI plot, and for M31\_5 and M31\_6 in the FeH plot, we show the best fit kinematic template spectra (coloured dashed lines) instead of the science data due to contamination from the sky background. Typical $\pm1\sigma$ error bars are denoted by the vertical lines at the bottom of each subplot.}
 \label{fig:spectra}
\end{figure*}

\subsection{Kinematic fitting and second-order sky subtraction}
\label{sec:secondsky}

We fit the kinematics of the spectra with the penalised pixel fitting routine \citep[{\sc ppxf}, ][]{CappellariEmsellem2004} using, as template spectra, the stellar population models from CvD12. Due to the presence of skyline residuals in the spectra, we found it necessary to perform a \emph{simultaneous} second-order sky subtraction while extracting the kinematics \citep{Weijmans2009}. This was accomplished using the techniques described in \citet{Davies2007}: rather than scaling a single sky spectrum covering multiple OH vibrational transitions, we divide the spectrum into separate regions covering each OH vibrational transition. This allows independent scaling of each transition. We also scale the O$_{2}$ emission around 0.864$\,\mu$m separately. Furthermore, to account for flexure (error in the wavelength calibration) we interpolate the spectrum of each transition (OH \& O$_{2}$) forward and backwards by one pixel. All these spectra are then passed to {\sc ppxf} which finds the best-fit linear summation that reproduces the skyline residuals while also finding the best fit kinematics\footnote{We manually altered the internal limits in {\sc ppxf} to allow negative sky spectrum weights}. For each galaxy spectrum (whether it be from a single bin or a median of the entire field-of-view), we measure the intrinsic instrument dispersion from the skylines in the same wavelength range. 

The SWIFT spectral resolution varies from $40\,\rmn{km}\,\rmn{s}^{-1}$ to $65\,\rmn{km}\,\rmn{s}^{-1}$. As the resolution of the CvD12 models is R=2000 beyond 0.75$\,\mu$m, it was necessary to correct the dispersions found by {\sc ppxf} for the difference in instrument and template resolution: to all {\sc ppxf} dispersions we added in quadrature the difference between the instrument resolution in that wavelength range and the template resolution. In practice this only significantly affects dispersions less than $100\,\rmn{km}\,\rmn{s}^{-1}$ (M31\_6 and M32). 

Using this technique, we are able to clean sky residuals from the galaxy spectra (prior to calculating element abundances) using the best-fit `second-order' sky spectrum generated by {\sc ppxf}. However, the sky spectrum contains both continuum and line emission, and scaling to remove the line residuals also scales the continuum. This may be a source of bias: adding or subtracting a constant continuum level from the galaxy spectrum will alter subsequent equivalent width measurements. As there is no reason to believe continuum and line emission scale in the same way \citep[unless the continuum is composed primarily of unresolved faint OH lines from the same transition,][]{Davies2007}, we subtracted the continuum from the second-order sky spectrum before using it to remove the skyline residuals in the galaxy spectra. Typically, the effect of this in equivalent width measurements is only a few per cent, but it can be as large as ten per cent.

For all fields, we fit the kinematics (and second-order sky spectra) separately for the NaI--CaT (0.785$\,\mu$m to 0.90$\,\mu$m) and FeH (0.97$\,\mu$m to 1.02$\,\mu$m) regions. For the FeH fit, we find it necessary to mask the region between 1.002$\,\mu$m and 1.008$\,\mu$m due to a significant template mismatch between the CvD12 and M31 spectra. For template spectra, we use only the solar metallicity SSPs covering a range of ages and IMFs. We do not include spectra for different element abundances as these are only available for a 13.5 Gyr Chabrier IMF SSP.

We tried masking the NaI and FeH absorption features during the kinematic fitting to test whether fitting the second-order sky spectrum affects the strength of these absorption features; no significant difference was found for data presented here.

\subsection{Index measurements}

Before making index measurements, we de-redshift each spectrum to correct for its velocity as determined from the kinematic fit, and convolve all our spectra up to a common resolution of $200\,\rmn{km}\,\rmn{s}^{-1}$ using a Gaussian of width given by
\begin{equation}
\sigma_{\rmn{conv}} = \sqrt{200^2 - (\sigma_*^2 + \sigma_{\rmn{instr}}^2)},
\end{equation}
where $\sigma_*$ in the stellar velocity dispersion measured by ppxf and $\sigma_{\rmn{instr}}$ is the instrumental resolution. Table~\ref{indices} gives the index and continuum definitions for the indices used in this paper.
\begin{table}
  \caption{Index bandpass and continuum definitions from \citet{Cenarro2001a} and \citet{ConroyVanDokkum2012a}. Wavelengths are in vacuum. We note several different definitions of NaI exist in the literature, including \citet{Vazdekis2012} and \citet{LaBarbera2013}.}
  \label{indices}
  \begin{tabular}{@{}lccc@{}}
  \hline
Index & Feature & Blue Continuum & Red Continuum\\
          & ($\rmn{\AA}$) & ($\rmn{\AA}$) & ($\rmn{\AA}$)\\
 \hline
 NaD & 5878.5-5911.0 & 5862.2-5877.2 & 5923.7-5949.7\\
 NaI & 8177.0-8205.0 & 8170.0-8177.0 & 8205.0-8215.0\\
 CaT & 8484.0-8513.0 & 8474.0-8484.0 & 8563.0-8577.0\\
         & 8522.0-8562.0 & 8474.0-8484.0 & 8563.0-8577.0\\
         & 8642.0-8682.0 & 8619.0-8642.0 & 8700.0-8725.0\\
 PaT & 8461.0-8474.0 & 8474.0-8484.0 & 8563.0-8577.0\\
         & 8577.0-8619.0 & 8563.0-8577.0 & 8619.0-8642.0\\
         & 8730.0-8772.0 & 8700.0-8725.0 & 8776.0-8792.0\\
 FeH & 9905.0-9935.0 & 9855.0-9880.0 & 9940.0-9970.0\\
\hline
\end{tabular}
\end{table}

The spectra around the NaI, CaT and FeH features are plotted in Fig.~\ref{fig:spectra}. For our resolved data of M31\_1 and M32, we plot a median spectrum from the entire field. We include the positions of the spectral index definitions from table~\ref{indices} in each subplot. Fig.~\ref{fig:spectra} shows the high quality of our measured spectra. Note that we have omitted the science spectra for M31\_6 in the NaI plot, and for M31\_5 and M31\_6 in the FeH plot. This is due to strong residual sky lines after optimisation during the sky subtraction stage discussed in section 2.4. We have instead plotted the best fit kinematic template spectra and we show a comparison between the data and templates in the appendix. We also note the definition of NaI appearing slightly blue-ward of the feature. The feature definition is over the Na doublet but there is a TiO band next to it, giving the appearance of a larger feature \citep{VanDokkumConroy2012}.

For M31\_1 and M32 we calculate radially binned average index values and standard errors on the mean from the individual spectra in the resolved fields. For M31\_2 to M31\_6 we combine the index uncertainty in quadrature with a conservative factor accounting for the uncertainty in telluric correction.

\begin{figure*}
 \centering
 \includegraphics[width=16.5cm]{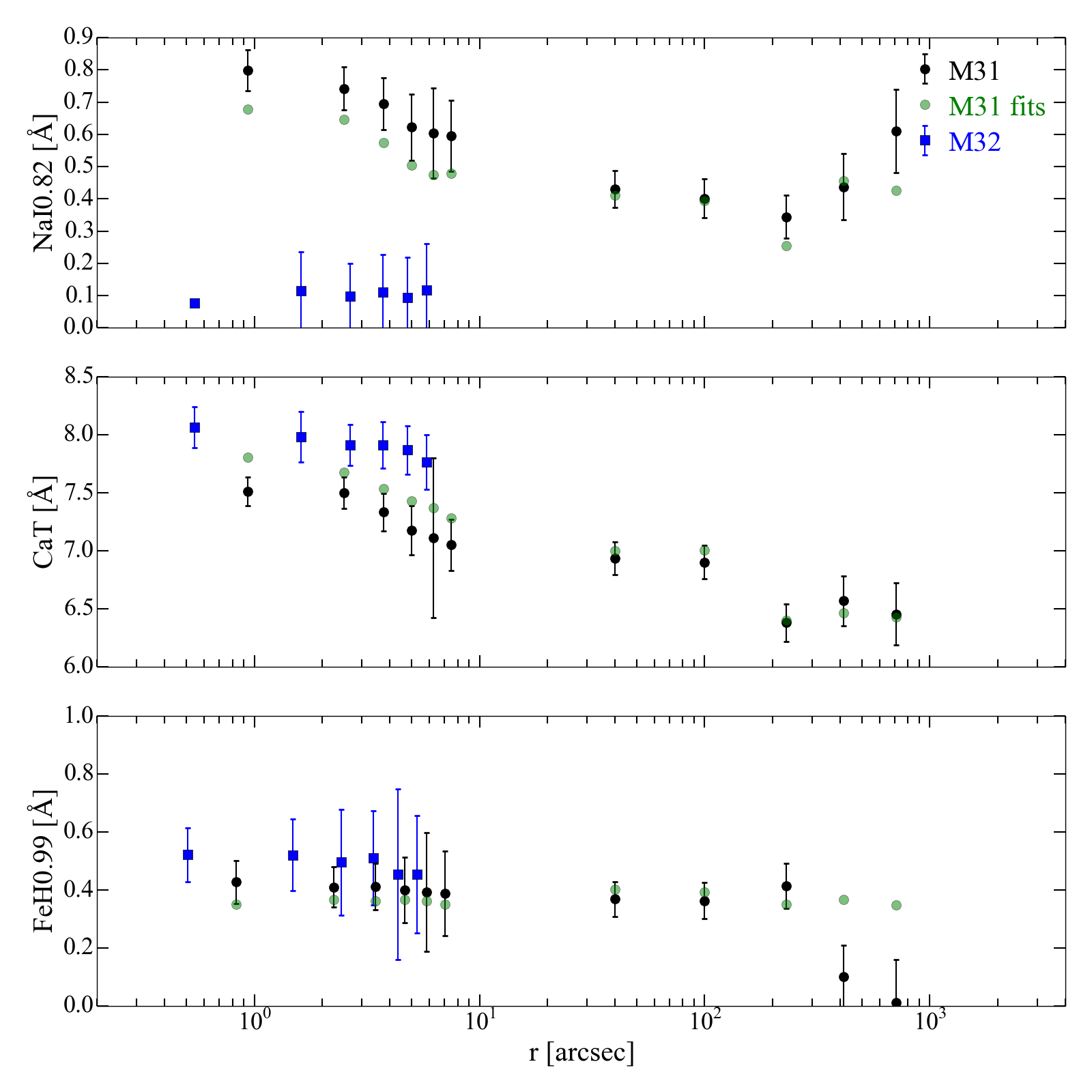}
 \caption{Plots of radial gradients for each index for M31 (black circles) and M32 (blue squares). Shown are the NaI, CaT and FeH indices as a function of radial distance. Our measurements of the best fit kinematic template spectra for M31 are shown as the green circles.}
 \label{fig:indicesvr}
\end{figure*}

\begin{figure*}
\begin{subfigure}
  \centering
  \includegraphics[width=8cm]{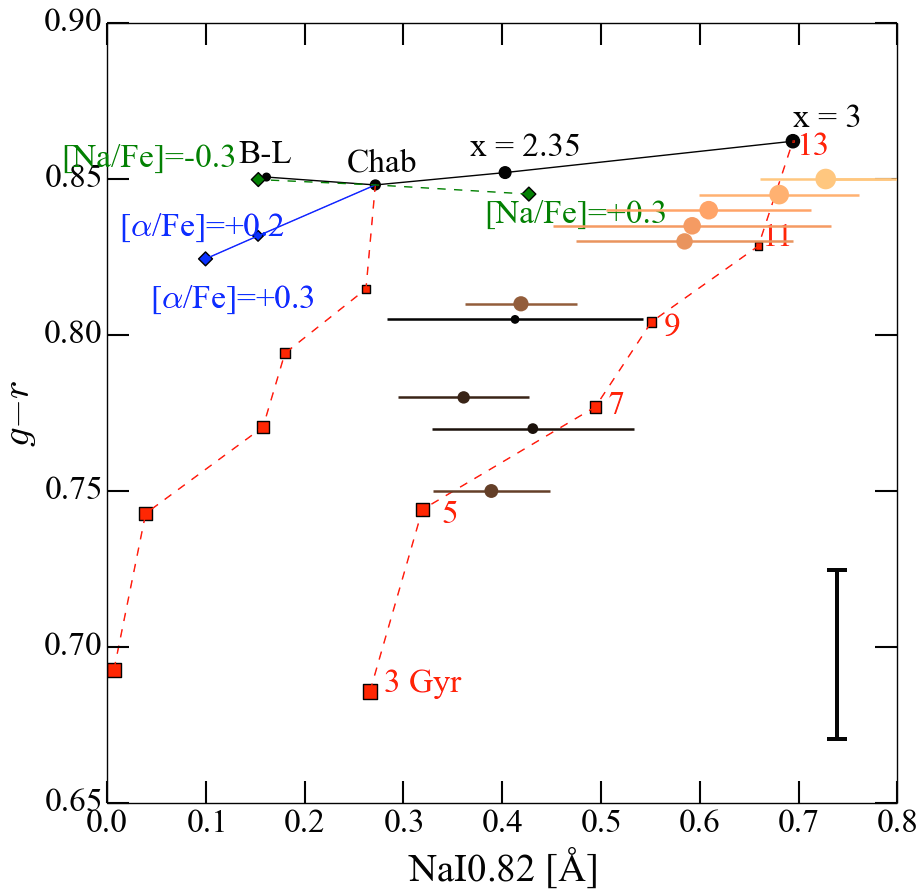}
  \label{fig:sfig1}
\end{subfigure}\begin{subfigure}
  \centering
  \includegraphics[width=9.35cm]{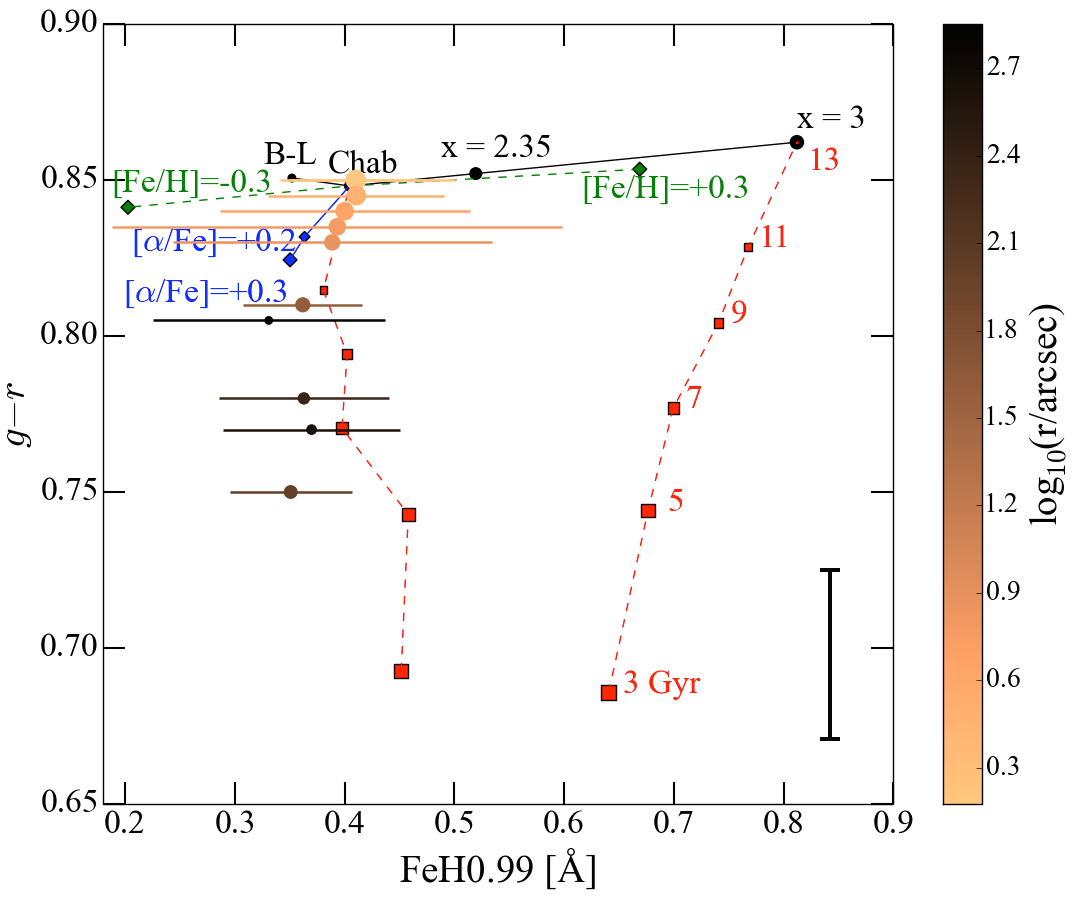}
  \label{fig:sfig2}
\end{subfigure}
\caption{Plots of NaI (left) and FeH (right) against SDSS $g-r$ colour showing CvD12 models with M31 data overplotted. B-L corresponds to a bottom-light IMF from \citet{VanDokkum2008}, and Chab to a \citet{Chabrier2003} IMF. The colour bar corresponds to the radial distance of each pointing to the centre of M31. Symbol sizes also decrease with increasing radial distance. Colour error bars are omitted from the points for clarity and are shown in the bottom right of each plot.}
\label{fig:NaI&FeHvcolour}
\end{figure*}

\section{results}

Our measurements of the three absorption features as a function of radius are shown in Fig.~\ref{fig:indicesvr} and we discuss each feature in turn for both galaxies.

The top panel in Fig.~\ref{fig:indicesvr} shows our measurements of the IMF sensitive index NaI. A clear negative gradient is visible in the central $\sim10''$ of M31, which becomes flat towards the edge of the bulge. Our measurement nearest the nucleus is a factor of two greater than the average for the bulge. In contrast, M32 displays a flat profile of low absorption. It is clear from Fig.~\ref{fig:indicesvr} that, for M31, our best fit template spectra have systematically lower values in the central $\sim10''$ compared to the observed spectra. However, in the outer bulge the two index measurements are in strong agreement (with the exception of M31\_6, which we show in the appendix). We believe the discrepancy in the central $10''$ originates from our input template spectra for {\sc ppxf} not including the SSPs with varying element abundances (as these spectra have a fixed age and IMF, see section \ref{sec:secondsky}).

CaT is shown in the second panel in Fig.~\ref{fig:indicesvr}. M31 displays a negative gradient throughout the bulge, dropping from $7.5\,\rmn{\AA}$ in the nuclear region down to $6.4\,\rmn{\AA}$ in the disc. The kinematic templates show a small systematic increase of $\sim0.25\,\rmn{\AA}$ compared to the measurements within $10''$, but are in very good agreement for the outer bulge measurements. The central discrepancy can be partly attributed to differences in the continuum measurements between the real and fitted spectra - the kinematic fits do not display as strong hydrogen absorption in the continuum. Residual telluric absorption could also be affecting the continuum level in the observed data. M32 displays very high calcium absorption of $\sim8.0\,\rmn{\AA}$ with a slight negative radial gradient.

Our measurements of FeH are shown in the bottom panel of Fig.~\ref{fig:indicesvr}. For both M31 and M32 FeH has a flat profile. The absorption strength is $0.4\,\rmn{\AA}$ for M31, and between $0.45-0.50\,\rmn{\AA}$ for M32.

\section{Analysis}

We analyse our measurements of the far red absorption features in M31 and M32 using the SSP models of CvD12 and V12. Both allow for variations in the IMF slope and age, however other parameters are specific to each set. CvD12 explore $\alpha$-element enhancement [$\alpha$/Fe], as well as variations in individual elemental abundance ratios [X/Fe] for 17 different elements (for fixed $13.5\,\rmn{Gyr}$ age and fixed Chabrier IMF). V12 explore variations in the total metallicity [Z/H] for all IMFs and ages. Note that the V12 models do not cover the FeH feature. Our spectra do not cover the commonly used age indicator H$\beta$, so we instead use SDSS $g-r$ colour. Fig.~\ref{fig:NaI&FeHvcolour} shows NaI (left) and FeH (right) against $g-r$ for the CvD12 models with the M31 data overplotted. The colour bar indicates the radial distance. Using these index maps, we now discuss our measurements in the context of our three goals: investigating the disc, bulge and nuclear region of M31, and the central region of M32. We show model variation plots for the V12 models in the appendix.

\begin{figure*}
 \centering
 \includegraphics[width=15cm]{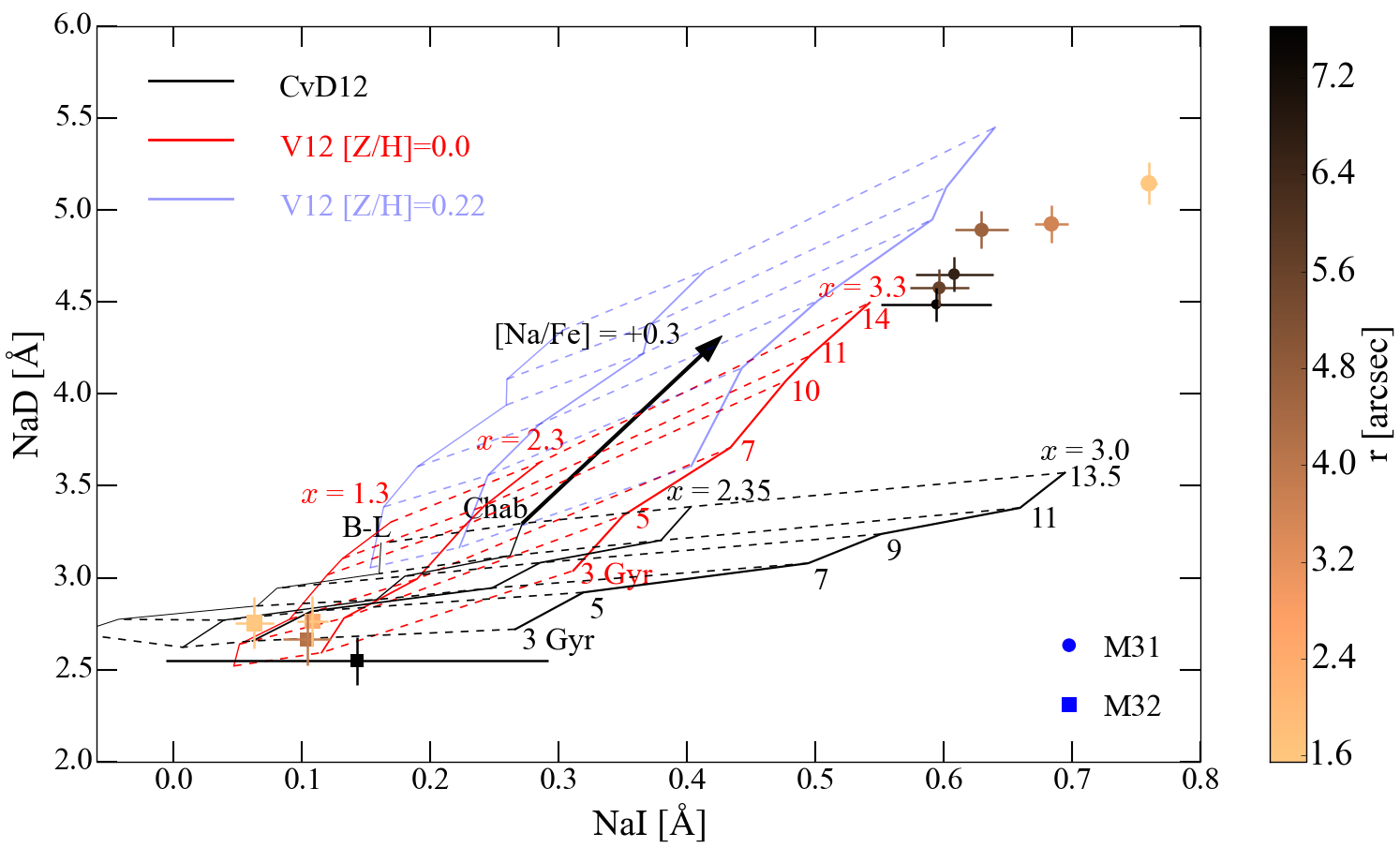}
 \caption{SSP model grids for CvD12 (black) and V12 (red and blue) showing variations in NaI against NaD. The CvD12 models are at solar metallicity, whereas the V12 models are at both solar (red, for M32) and metal enhanced with [Z/H]=+0.22 (blue, for M31). Note that the varying age vectors from the V12 models are parallel to the [Na/Fe] vector from the CvD12 models (black arrow). The central $10''$ of M31 (circles) and M32 (squares) are plotted. Symbol sizes decrease with increasing radial distance. We use the NaD measurements of \citet{Davidge1991, Davidge1997} that cover $r\leq10''$. IMF labels are the same as in Fig.~\ref{fig:NaI&FeHvcolour}.}
 \label{nai_v_nad}
\end{figure*}

\subsection{M31}

Fig.~\ref{fig:NaI&FeHvcolour} shows NaI (left) and FeH (right) index measurements plotted against SDSS $g-r$ colour as a function of radius. We over plot our data onto CvD12 model predictions for both indices. NaI appears to suggest an IMF slope around $x=3$ for the central $10''$ with the bulge values also in a region of steepened IMF slope. In contrast, the FeH data follow the CvD12 predictions for a Chabrier IMF at solar metallicity closely. The discrepancy between NaI and FeH predictions can be explained by Na enhancement within the bulge of M31. The left plot in Fig.~\ref{fig:NaI&FeHvcolour} shows that a Chabrier IMF with Na enhancement of [Na/Fe]$\, =+0.3\,\rmn{dex}$ could cover the outer bulge measurements. The inner bulge and nuclear region would require even stronger Na enhancement, with linear extrapolation of the models suggesting up to an order of magnitude above solar. The $g-r$ data indicate a decreasing age between inner and outer bulge along the major axis, although the large errors of $\pm0.03\,\rmn{mag}$ do not allow for any stronger conclusions. However, this is in general agreement with best fit ages derived by \citet{Saglia2010}.

CaT is known to be inversely proportional to IMF slope \citep[see e.g.][]{Cenarro2003}. However, it is much more strongly influenced by [$\alpha$/Fe] and Ca abundance than by IMF. The CvD12 models suggest an order of magnitude increase in the sensitivity of CaT to changes in [$\alpha$/Fe] compared to changes in IMF (see Fig.~\ref{A:V12models}). The gradient we measure is indicative of a steepening IMF slope with radius. Furthermore, this is in contrast to the FeH measurements. CvD12 note that Na enhancement can change the CaT strength. Thus, the flat profile of FeH leads us to conclude that the bulge of M31 holds a Chabrier IMF and is Na enhanced throughout, more so in the central $10''$.

Fig.~\ref{nai_v_nad} shows the central $10''$ of M31 on an index-index plot of the two sodium features, NaI and NaD. Model grids for CvD12 are fixed at solar metallicity (black), whereas those of V12 are at both solar (red) for M32 \citep{Rose2005} and [Z/H]$\,= +0.22$ (blue) for M31 \citep{Saglia2010}. The radial measurements of the M31 inner bulge and nuclear region follow the V12 model lines of varying age at a very bottom-heavy $x>3.3$ IMF exponent. Thus the V12 models suggest a bottom-heavy IMF in this region, which would become Salpeter-like through the rest of the bulge as NaI approaches $0.4\,\rmn{\AA}$. However, also plotted is a CvD12 SSP at $13.5\,\rmn{Gyr}$, with a Chabrier IMF and sodium enhanced by $+0.3\,\rmn{dex}$. It is clear from this SSP that the CvD12 models attribute a stronger dependence of NaD on [Na/Fe] than the IMF slope, and further it is indicative that the CvD12 models suggest the level of Na enhancement seen in the central $10''$ of M31 is consistent with a normal (between Chabrier and Salpeter) IMF. It is not possible to resolve the differences between the two model predictions without coverage of the FeH feature and elemental abundance variations in the V12 models.

\subsection{M32}

M32 displays flat profiles for both NaI and FeH, with a slightly negative gradient in CaT. NaI is markedly lower than in M31 with values around $0.1\,\rmn{\AA}$. From Fig~\ref{nai_v_nad} both the CvD12 and V12 models suggest this is consistent with a Chabrier IMF at solar metallicity around 3--5 Gyr in age. The stellar populations of M32 are younger than M31 \citep[see e.g. ][]{OConnell1980} and the model predictions for the age from both CvD12 and V12 are in reasonable agreement with long-slit spectroscopy measurements by \citet{Rose2005}, who measure an age gradient from 3--4$\,\rmn{Gyr}$ in the centre to 6--7$\,\rmn{Gyr}$ at $30''$. A younger stellar population with Chabrier IMF is further supported by the FeH index, which is slightly stronger than in M31.

A striking difference is the much higher CaT values compared to M31. This has previously been noted by \citet{KormendyBender1999} and \citet{Cenarro2008}. \citet{Thomas2003} suggest this can be explained by the fact that M32 is younger than M31, and the younger stars contaminate the CaT index, which is known to be affected by Paschen lines. However, the CaT* index is defined by \citet{Cenarro2001a} as CaT* $= \rmn{CaT}-0.93\times\rmn{PaT}$, where PaT is an index defined over several Paschen lines next to CaT, in order to calibrate out the hydrogen lines. We measured the CaT* index within the central $2''$ of M32 as CaT* $=7.39\pm0.22\,\rmn{\AA}$ and within the same aperture of M31 as CaT* $=6.75\pm0.16\,\rmn{\AA}$, which still shows greater absorption for M32. It would be an interesting test to see whether other dwarf galaxies of the same age as M32 also display strong calcium absorption. Overall our measurements of M32 are fully consistent with a Chabrier IMF.

\section{Discussion}

The form of the IMF in the nuclear region of M31 has been of intense historical debate. The main factor driving the idea of a bottom heavy IMF has been the prominent NaI feature in the central $\sim3''$. \citet{SpinradTaylor1971} interpreted this as the presence of a large number of dwarf stars. Further observations by many groups \citep[see e.g.][]{Whitford1977, Frogel1978, Cohen1978, FaberFrench1980, Carter1986, AlloinBica1989, DelisleHardy1992} have led to conflicting results. \citet{ConroyVanDokkum2012b} observed the central 15 pc of M31 ($4''$) and inferred a mass-to-light ratio consistent with an IMF slope between Chabrier and Salpeter. They concluded that the central region is entirely consistent with a Milky Way IMF, albeit requiring very large sodium enhancement of the order [Na/Fe]$\sim1.0$.

All current work suggesting increasing IMF slopes with velocity dispersion in the centres of galaxies has been for massive early-types, which generally have large dispersions, upwards of $300\,\rmn{km}\,\rmn{s}^{-1}$. \citet{Ferreras2013}, \citet{LaBarbera2013} and \citet{Spiniello2014b} find similar empirical relations between the IMF slope and velocity dispersion. M31 has a dispersion of $\sigma\sim160\,\rmn{km}\,{s}^{-1}$ (out to $0.4\,\rmn{kpc}$ which we adopt as a typical central dispersion), which gives an IMF slope of $x\sim1.9$ from all three relations and is consistent with a normal (i.e. not bottom-heavy) IMF. We measure a Chabrier IMF throughout M31 even at the largest radius which has a dispersion of $\sigma\sim90\,\rmn{km}\,\rmn{s}^{-1}$, and also in the centre of M32 which has a central dispersion of $\sim60\,\rmn{km}\,\rmn{s}^{-1}$; this suggests that the relations linking IMF to velocity dispersion should become fixed at Chabrier IMF for $\sigma<150\,\rmn{km}\,\rmn{s}^{-1}$. \citet{Chabrier2014} have presented a theoretical scenario for dwarf enrichment and a steepening of the IMF within the framework of gravitationally collapsing turbulent molecular clouds. They deduce an upper limit to the IMF slope of $x\sim2.8$ in extremely high density and turbulent systems. We note that the centres of M31 and M32 have very high stellar density and yet we measure a Chabrier IMF.

The physical size scales here are also worth considering. The central $8''$ of M31 corresponds to 30 pc, which is a very small region of space in which to create and maintain an overabundance of dwarf stars (or in fact, a population with [Na/Fe]$\sim+1.0$). Furthermore, sodium excess in stars \citep{Briley1995} and in galaxies \citep{Jeong2013} is not particularly rare, suggesting that M31 may not be peculiar. In fact, \citet{Jeong2013} measure NaD absorption as high as $\sim7\,\rmn{\AA}$ in some ETGs and late type galaxies (LTGs); much higher than seen in M31. They conclude that the NaD index in most Na enhanced LTGs is strongly contaminated by the interstellar medium (ISM) and/or dust. These findings caution against using sodium as a sole IMF indicator, especially for late type galaxies. Most importantly, our measurements of the FeH absorption reject the possibility of a dwarf-dominated nuclear region. \citet{ConroyVanDokkum2012b} also reached this conclusion for the central $4\rmn{''}$ and our results have extended that conclusion throughout the bulge and into the disc.

It is an important point that the two SSP models give different predictions when considering the sodium indices as highlighted by \citet{Spiniello2015}. Specifically V12 attribute a much stronger dependence of NaD on the IMF slope than the CvD12 models. This is clear in Fig.~\ref{nai_v_nad} and gives apparently discrepant predictions for the IMF slope in M31. We conclude with \citet{Spiniello2015} that sodium is complex and poorly understood. However, the V12 models do not cover the FeH wavelength range so no constraint on the IMF can be made using this feature with the V12 models.

\section{Conclusions}

Using the Oxford SWIFT integral field spectrograph we have undertaken a study of M31 and M32 with the aim of discerning possible gradients in IMF-sensitive far red absorption features. We obtained high S/N integral field spectroscopy of M31, covering the nuclear region, bulge and old disc at $\sim700'' (2.7\,\rmn{kpc})$, as well as for the central $10'' (38\,\rmn{pc})$ of M32. We conclude the following:

1. In M31, the NaI index shows a flat equivalent width profile at $\sim0.4\,\rmn{\AA}$ through the majority of the bulge, with a strong gradient up to $0.8\,\rmn{\AA}$ in the central $10''$. FeH is measured to be flat at $0.4\,\rmn{\AA}$ on all scales. CaT shows a gradual positive gradient inwards through the bulge. M32 displays flat profiles for all three indices: slightly stronger FeH absorption than M31, anomalously strong CaT absorption in agreement with other studies, and very weak NaI absorption.

2. M31 is consistent with a Chabrier IMF throughout the central region, bulge and old disc from analysis of the flat FeH index using the CvD12 models. We also infer a gradient in sodium enhancement of [Na/Fe] $\sim+0.3\,\rmn{dex}$ in the disc and outer bulge, rising up to [Na/Fe] $\sim+1.0\,\rmn{dex}$ in the nuclear region.

3. Our measurements of NaI and FeH for M32 are consistent with a Chabrier IMF for ages $\sim$\,3--5$\,\rmn{Gyr}$ using the CvD12 and V12 models.

4. The use of sodium indices alone is a poor method for investigating IMF variations as it is susceptible to sodium enhancement and [$\alpha$/Fe] variations. Sodium is a poorly understood element as highlighted by the discrepant predictions of the CvD12 and V12 models.

5. The centres of M31 and M32 have very high stellar densities and yet we measure a Chabrier IMF in both regions.

\section*{Acknowledgments}

We are grateful to the anonymous referee for providing feedback that significantly improved this paper. The Oxford SWIFT integral field spectrograph was supported by a Marie Curie Excellence Grant from the European Commission (MEXT-CT-2003-002792, Team Leader: N. Thatte). It was also supported by additional funds from the University of Oxford Physics Department and the John Fell OUP Research Fund. Additional funds to host and support SWIFT at the 200-inch Hale Telescope on Palomar were provided by Caltech Optical Observatories.

The authors thank C. Conroy for kindly providing SSP models and T. Davidge for kindly providing NaD data for M31. We are also grateful to F. La Barbera, I. Ferreras and A. Vazdekis for useful feedback and for clarifying that the V12 models are defined in air wavelengths. This work was supported by the Astrophysics at Oxford grants (ST/H002456/1 and ST/K00106X/1) as well as visitors grant (ST/H504862/1) from the UK Science and Technology Facilities Council. SZ is supported by STFC-HARMONI grant ST/J002216/1. RCWH was supported by the Science and Technology Facilities Council [STFC grant numbers ST/H002456/1 \& ST/K00106X/1]. RLD acknowledges travel and computer grants from Christ Church, Oxford, and support for a visit from the Space Telescope Science Institute in the final stages of this work.

\bibliography{./MASTERBIB}
\bibliographystyle{mn2e_new}

\appendix

\section[]{Additional plots}

We show here several additional plots concerning the accuracy and limitations of our telluric correction and sky subtraction techniques, as well as comparison plots between CvD12 and V12 models.

\begin{figure*}
\begin{subfigure}
  \centering
  \includegraphics[width=8.3cm]{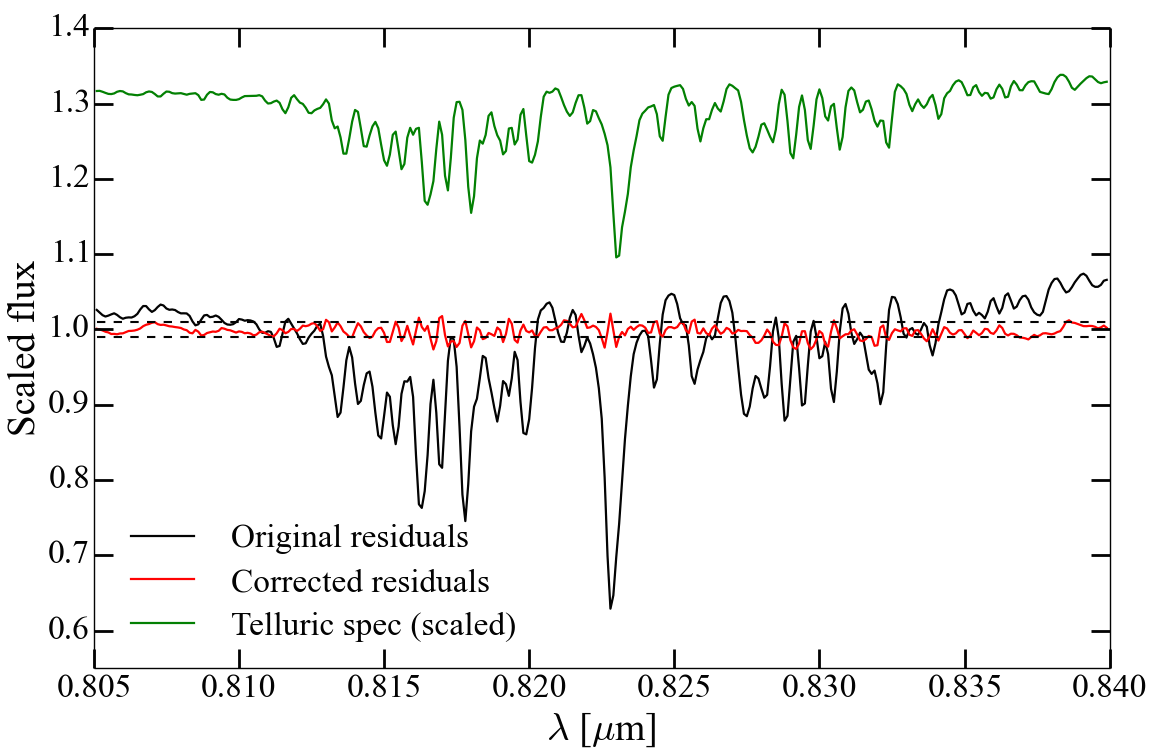}
  \label{fig:sfig1}
\end{subfigure}\begin{subfigure}
  \centering
  \includegraphics[width=8.3cm]{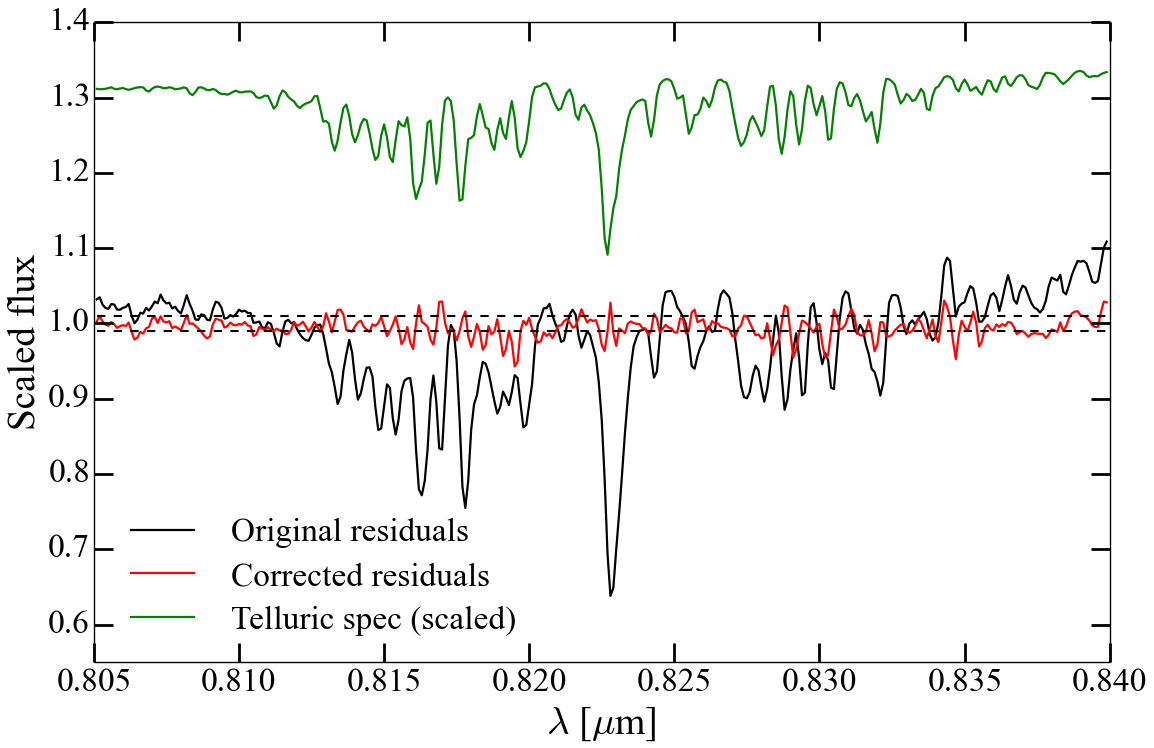}
  \label{fig:sfig2}
\end{subfigure}
\caption{Telluric correction for M31\_2 (left) and M31\_6 (right). The black line shows the ratio between original spectrum and kinematic template (assuming this is the intrinsic `telluric free' spectrum), the red line shows the same ratio with telluric division, and the green line shows the telluric spectrum scaled by factor of 0.5 and plotted at 1.3 for clarity.}
\label{telluric_correction}
\end{figure*}

Fig.~\ref{telluric_correction} shows examples of our telluric correction around the NaI feature for M31\_2 and M31\_6. In each plot we use our kinematic template fits as an assumed `telluric free' galaxy spectrum. The residuals after telluric correction are 0.8 per cent and 1 per cent rms respectively. We also check this for CaT and FeH and find residuals $< 1$ per cent for these features.

\begin{figure}
 \centering
 \includegraphics[width=7.5cm]{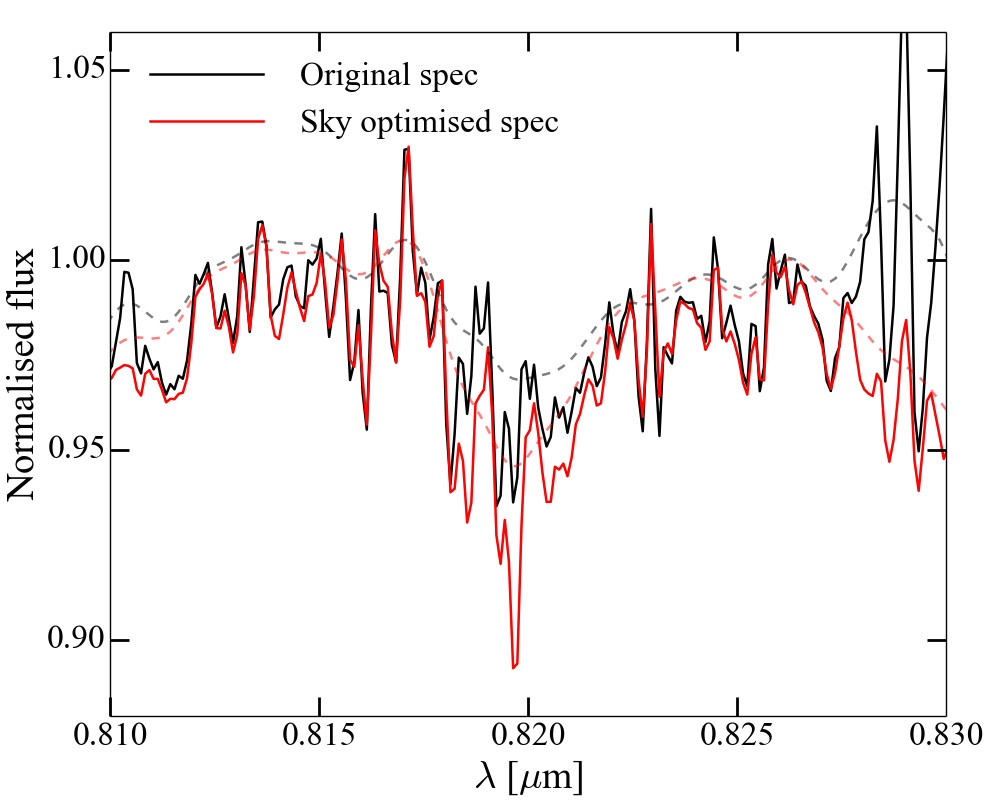}
 \caption{Plot showing the M31\_6 spectrum before (black) and after (red) the 2nd order sky subtraction routine (section 2.4). The procedure has deepened the NaI feature with (false) excess sodium emission, which leads to a larger index value from the convolved spectra (dashed lines).}
 \label{A:m31_6_original_v_optimsed}
\end{figure}

Fig.~\ref{A:m31_6_original_v_optimsed} shows the M31\_6 spectrum around the NaI feature. The second-order sky subtraction procedure has deepened the feature with extra sodium absorption, leading to a deeper feature. We checked this for all spectra to ensure systematic errors were minimised and only found this contamination in the M31\_6 spectrum. We show this plot to highlight why we have used the best fit kinematic template to measure NaI for M31\_6 instead of the science spectrum.

\begin{figure}
 \centering
 \includegraphics[width=7.5cm]{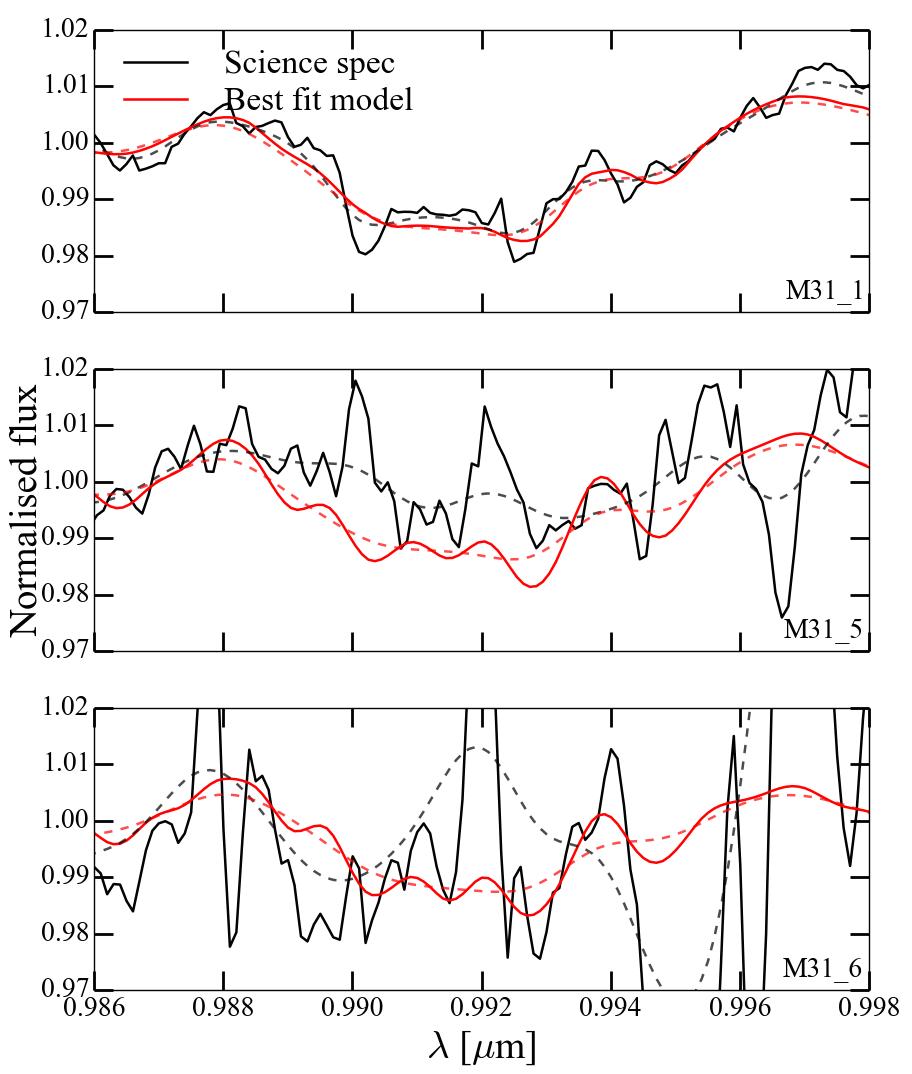}
 \caption{Plotted spectra showing science spectra (black lines) with best fit models (red lines) for three fields, M31\_1, M31\_5 and M31\_6. Solid lines show the spectra at their intrinsic resolution, and dashed lines show the spectra convolved up to $200\,\rmn{km}\,{s}^{-1}$. The spectra have each been normalised over the pseudo-continuum of the FeH feature.}
 \label{A:m31_5-m31_6}
\end{figure}

Fig.~\ref{A:m31_5-m31_6} shows our M31\_1, M31\_5 and M31\_6 science spectra around the FeH features with their best fit kinematic templates. The presence of residual sky lines after the second order sky subtraction stage in the M31\_5 and M31\_6 spectra leads us to use the template spectra to measure the FeH feature for these fields instead of the science spectra.

\begin{figure*}
\begin{subfigure}
  \centering
  \includegraphics[width=7.8cm]{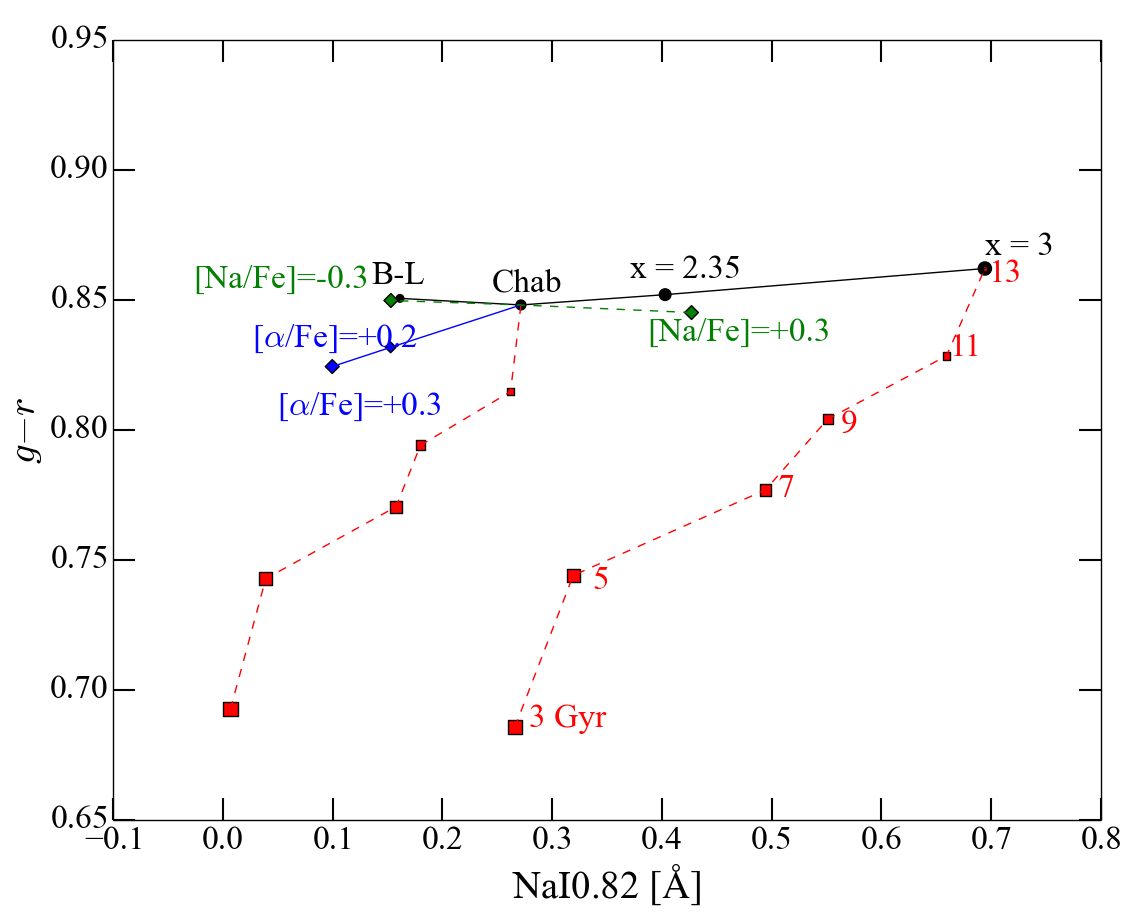}
  \label{fig:sfig1}
\end{subfigure}
\begin{subfigure}
  \centering
  \includegraphics[width=7.8cm]{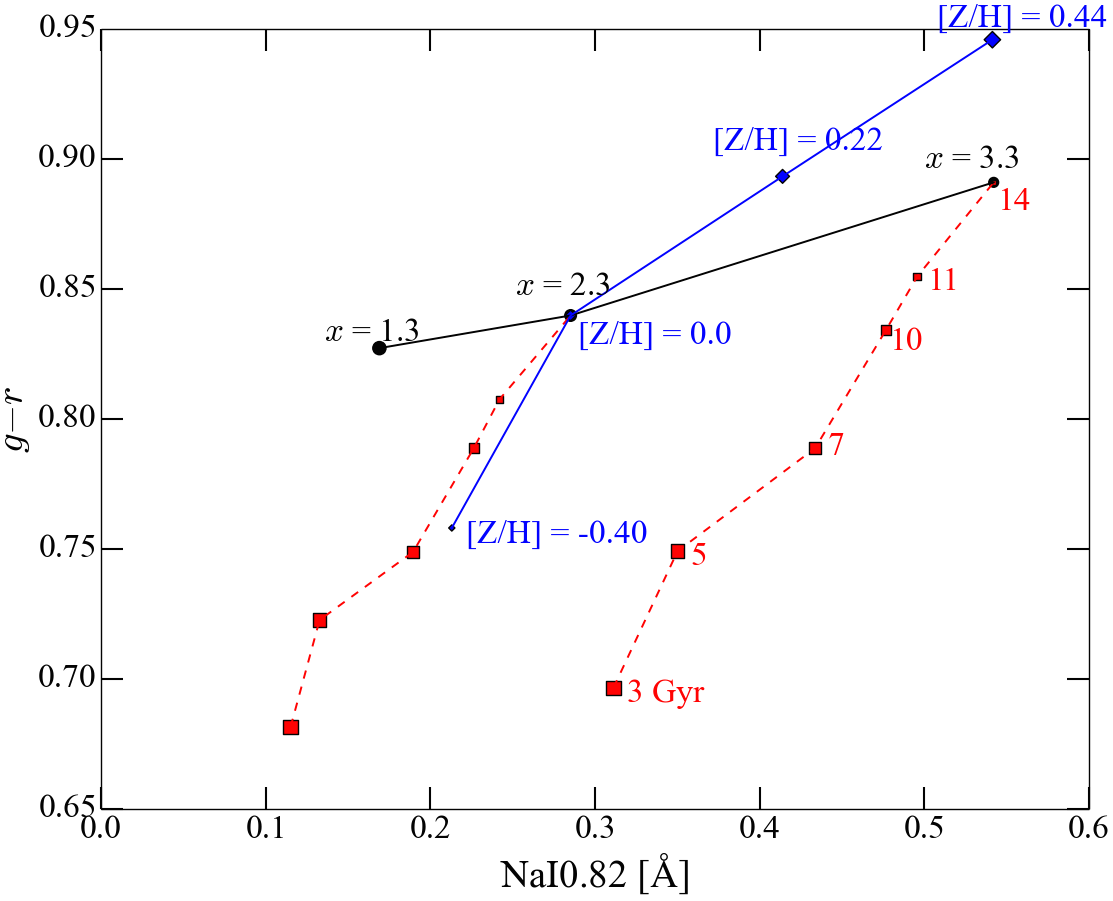}
  \label{A:sfig2}
\end{subfigure}\begin{subfigure}
  \centering
  \includegraphics[width=7.8cm]{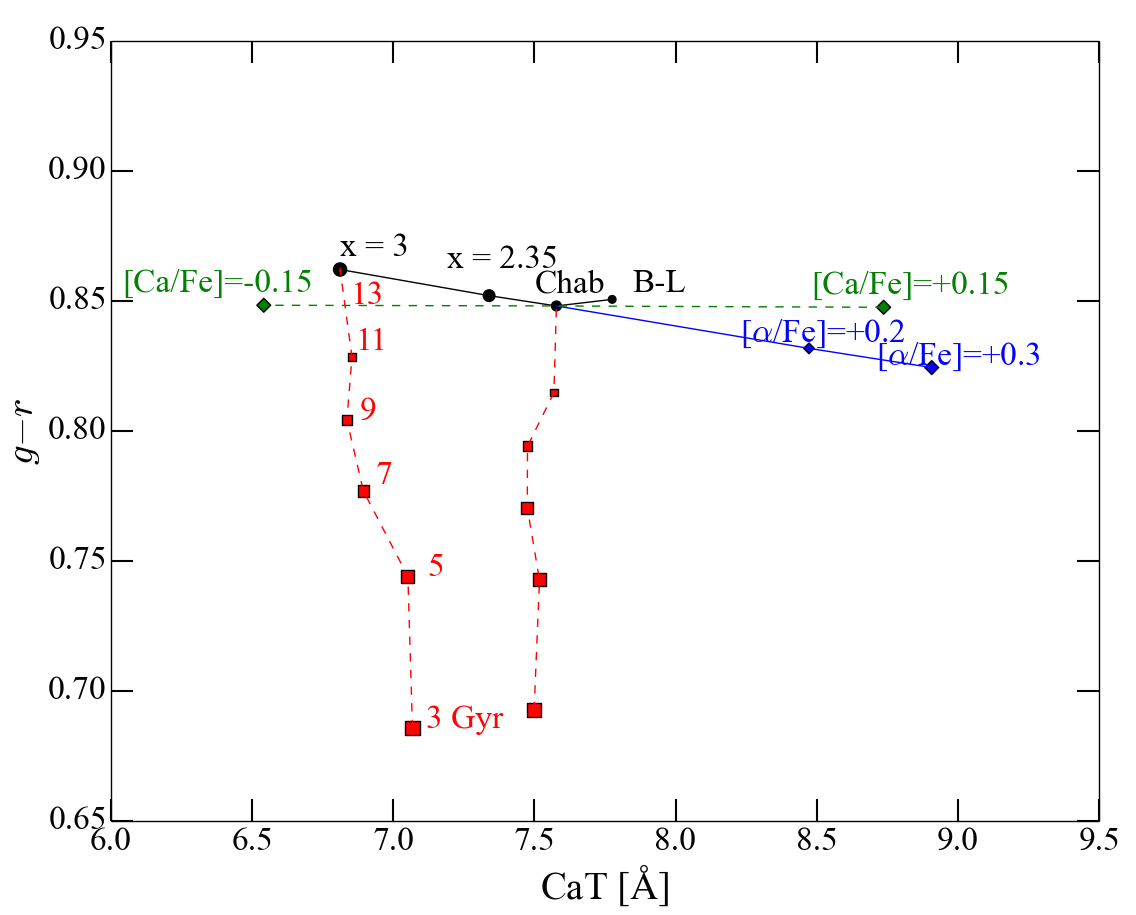}
  \label{fig:sfig3}
\end{subfigure}
\begin{subfigure}
  \centering
  \includegraphics[width=7.8cm]{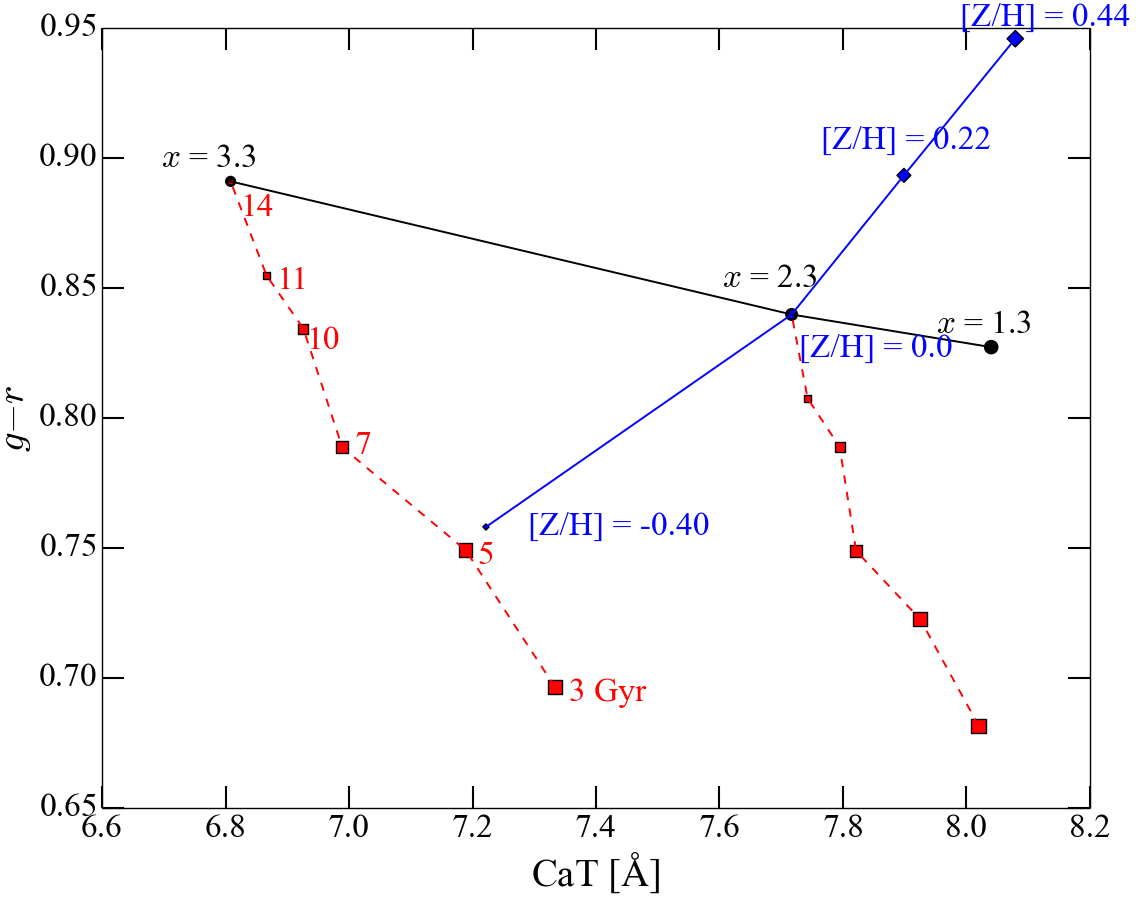}
  \label{fig:sfig4}
\end{subfigure}\begin{subfigure}
  \centering
  \includegraphics[width=7.8cm]{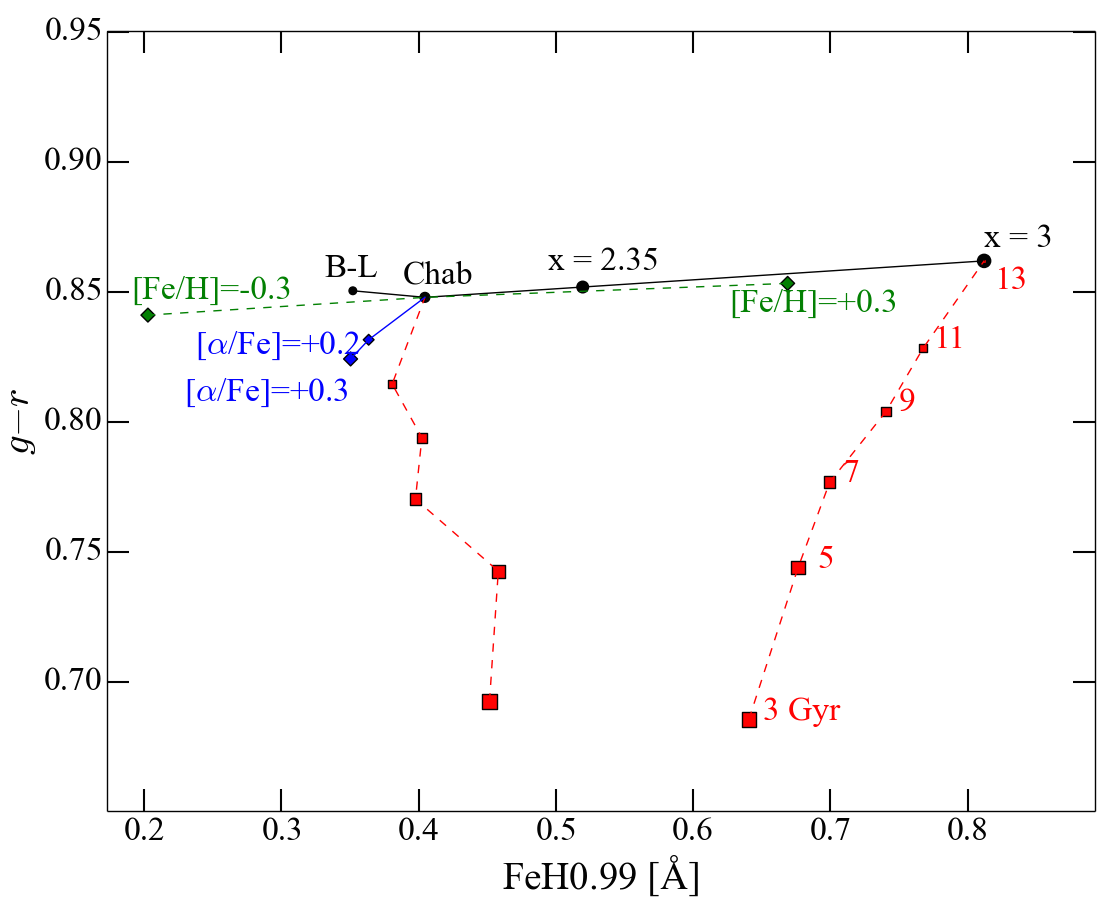}
  \label{fig:sfig5}
\end{subfigure}
\caption{Plots of NaI index (top row) and CaT (second row) against SDSS $g-r$ colour showing model variation for CvD12 (left) and V12 (right) models. Also shown is FeH (bottom) against $g-r$ for CvD12 models only, as FeH is not covered by V12. For the V12 models, black lines and circles are different IMFs at 14 Gyrs and [Z/H]=0.0, red squares are different ages, and blue diamonds are different metallicities. For the CvD12 models, black lines and circles show different IMFs, red lines and squares show different ages, blue lines and diamonds show varying [$\alpha$/Fe] abundance, and green dashed lines and diamonds show varying [X/Fe].}
\label{A:V12models}
\end{figure*}

Fig.~\ref{A:V12models} shows comparison model grids between CvD12 (left) and V12 (right) for NaI (top row), CaT (second row) and FeH (bottom; CvD12 only) features. We plot each index against variation in SDSS $g-r$ colour. These maps serve to show and compare the behaviour of each set of SSP models for each index within their respective model parameter spaces. Both sets of models qualitatively agree in their behaviour for variations in IMF and age. However quantitative differences exist, especially for the CaT index.

\bsp

\label{lastpage}

\end{document}